\documentclass[prl,twocolumn,showpacs,amsmath,amssymb]{revtex4}

\input epsf
\usepackage{graphicx}
\usepackage{dcolumn}

\usepackage{bm}

\begin{document}

\preprint{}

\title{X-ray observation of micro-failures in granular piles approaching an avalanche}

\author{Alexandre Kabla$^1$}
\author{Georges Debr\'egeas$^2$}
\author{Jean-Marc di Meglio$^{1,3}$}
\author{Tim J. Senden$^1$}
\email{Tim.Senden@anu.edu.au}
 \affiliation{{$^1$Department of Applied Mathematics, Research School of Physical Sciences and
Engineering, The Australian National University, Canberra,
Australia}\\
 {$^2$LFO-Coll\`ege de France, CNRS UMR 7125, Paris,
 France}\\
 {$^3$MSC, CNRS, Universit\'e Paris 7, Paris, France}}
\date{\today}

\begin{abstract}

An X-ray imaging technique is used to probe the stability of
3-dimensional granular packs in a slowly rotating drum. Well
before the surface reaches the avalanche angle, we observe
intermittent plastic events associated with collective
rearrangements of the grains located in the vicinity of the free
surface. The energy released by these discrete events grows as the
system approaches the avalanche threshold. By testing various
preparation methods, we show that the pre-avalanche dynamics is
not solely controlled by the difference between the free surface
inclination and the avalanche angle. As a consequence, the measure
of the pre-avalanche dynamics is unlikely to serve as a tool for
predicting macroscopic avalanches.
\end{abstract}

\pacs{45.70.Cc, 45.70.Ht, 81.05.Rm}

\maketitle

When a granular pack is submitted to a slowly varying stress its
apparent response consists in intermittent bursts of plasticity
during which large irreversible deformations take place
\cite{Nasuno1997,Howell1999}. Between these discrete events the
system behaves as a rigid body. Formation of shear-bands in
granular systems in triaxial tests \cite{Besuelle2000}, the
jamming and unjamming of an hourglass \cite{Bertho2003} or
sequences of avalanches down a sandy slope are common examples of
this intermittent behavior. Because of its relevance to geophysics
and the possibility of direct observations, the case of avalanches
have been studied in great detail since the early work of Bagnold
\cite{Bagnold}. It was found that the intermittent regime occurs
between two limiting angles of the heap surface: the repose and
the avalanche angle which depend on the geometry as well as the
characteristics of the material (shape and surface properties of
the grains) \cite{Duran2000}. Although the main features of the
avalanche flow are rather well understood
\cite{Daerr1999,Rajchenbach2002,GdRMidi2004}, one important
question remains to date unsolved: what is the nature of the
microscopic process by which the flow is triggered and later
stopped during a single avalanche?

To address this question several experimental and numerical
studies have recently focused on micro-plasticity
\cite{Staron2002,Deboeuf2003}. Indeed it appears that even in the
absence of visible flow, any modification of the external force
applied to a granular pack induces some micro-displacements which
allow the system to mechanically adapt to the new constraints. By
analogy with earthquakes, these events could be seen as precursors
of the macroscopic failure of the pack. Although they dissipate
only a small fraction of the energy as compared to the macroscopic
avalanches, they could impact the stress distribution within the
static pack therefore controlling its
overall stability.\\

\begin{figure}
\centerline{ \epsfxsize=8truecm \epsfbox{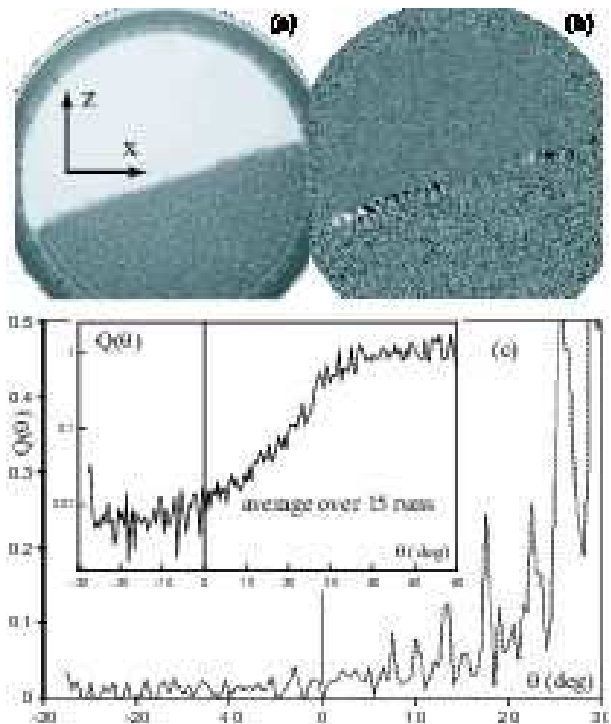}}
\caption{Measurement of the dynamics induced by a tilt increment
for the system I (see Table \ref{systems}): (a) radiography of the
drum, (b) picture presenting the variations of the crossed grains
length $\Delta L_g$, at an angle 20$^\circ$, (c) measurements of
the quadratic fluctuations as a function of the drum angular
position $\theta$. The doted line corresponds to a single run,
whereas the inset (semi-log scale) shows the average over 15 runs.
The data have been rescaled by the average of $Q$ in the avalanche
regime.}
 \label{figure1}
 \end{figure}

In the present work, we use an X-ray imaging technique to probe
the energy and spatial characteristics of these precursors in 3D
granular systems slowly tilted towards the avalanche angle.
Experiments are done in a cylindrical acrylic drum $55$ mm in
diameter with a depth of $30$ mm. The drum is half filled with
various types of grains, Table \ref{systems}, and mounted with its
rotation axis parallel to the X-ray beam. To prevent a bulk
rotation of the sample, a disordered monolayer of grains is glued
to the inner surface of the cylinder. The drum is co-axially
mounted inside the central aperture of a precision rotation stage
which allows for very smooth rotation and a maximal rotation speed
of 0.5$^\circ$s$^{-1}$. A micro-focus X-ray source is placed 230
cm from the drum and provides a weakly divergent beam which
crosses the stage aperture, producing a radiograph on a
scintillator-coupled CCD camera \cite{Sakellariou2004}, figure
\ref{figure1}(a). For a monochromatic beam, the transmitted
intensity of the incident beam at a position $[x,z]$ reads:
\begin{equation}\label{IntensityMap}
\textbf{I}[x,z] = A(t) \cdot \textbf{I$_0$}[x,z] \cdot
\textbf{a$_c$}[x,z] \cdot \exp \left( -
\frac{\textbf{L$_g$}[x,z]}{\lambda_g} \right)
\end{equation}
In this expression, $A(t) \cdot \textbf{I}_0[x,z]$ is the incident
intensity map without sample ($A(t)$ being the temporal fluctuations
of the X-ray source intensity), $\textbf{a}_c[x,z]$ corresponds to
the absorption of the empty container, and $\lambda_g$ is the
absorption length of the grain material. The total length of the
grain material which projects through each position $[x,z]$ on the
drum is denoted, $\textbf{L}_g[x,z]$.

\begin{table}
  \centering
\begin{tabular}{|c|c|c|c|c|}
 \hline
  $\;\;$   $\;\;$& $\;\;$ Material $\;\;$& $\;\;$ Shape  $\;\;$& $\;\;$ Size  $\;\;$ & $\;\;$$\theta_a$$\;\;$ \\ \hline
  I & Pasta & $\;\;$ cylinder  $\;\;$& $\;\;$ 2.2mm  $\;\;$ & 30 \\
  II & Carbon & sphere & 1.1mm & 27 \\
  $\;\;$III$\;\;$ & Carbon & sphere & 0.8mm  & 26\\
  IV & Carbon & sphere & 0.4mm & 22\\ \hline
\end{tabular}\\
  \caption{Different types of grains used in the experiments.
$\theta_a$ is defined as the average inclination angle of the
surface in the avalanche regime. The size corresponds to the
diameter for the spheres. The cylinder is $2.2\pm 0.2$mm in height
and diameter. Glassy carbon beads were obtained from
Hochtemperatur-Werkstoffe Gmbh, pasta was De Cecco Acini di
pepe.}\label{systems}
\end{table}

To prepare the grain pack in a reproducible way, the drum is
initially rotated 360$^\circ$ clockwise, leaving the pack surface at
the avalanche angle  $-\theta_a$.  The drum is then rotated
incrementally counter-clockwise.  To probe whether a displacement
has occurred over any given increment  $\delta \theta$
(0.5$^\circ$), we use the following protocol: at $\theta$ a
radiograph of the pack is taken, $\textbf{I}_1[x,z](\theta)$. The
drum is further rotated to the angle $\theta + \delta \theta$,
stopped for 5 s, then rotated back to $\theta$ whereupon another
radiograph $\textbf{I}_2[x,z](\theta)$ is taken. Finally, the drum
rotates back to $\theta + \delta \theta$ in preparation for the next
increment. The change in $\textbf{L}_g[x,z]$, by direct comparison
between $\textbf{I}_1[x,z](\theta)$ and $\textbf{I}_2[x,z](\theta)$
quantitatively characterizes the pack dynamics:

\begin{equation}\label{DLgMap}
\Delta \textbf{L}_g(\theta) = \lambda_g \cdot \left[
 \ln \left( \frac{\textbf{I}_1(\theta)}{\textbf{I}_2(\theta)} \right) -
 \left\langle \ln  \left( \frac{\textbf{I}_1(\theta)}{\textbf{I}_2(\theta)}
 \right)\right\rangle_{[x,z]} \right]
\end{equation}

The average in eq. \ref{DLgMap} corrects for temporal fluctuations
in the X-ray source. A grey-scale representation of $\Delta
\textbf{L}_g[x,z]$ is presented in figure \ref{figure1}(b). This
image demonstrates the existence of movements in the vicinity of
the free surface. A closer examination reveals that these motions
occur in the form of clusters of a few grains, moving over a
fraction of the grain size.

Several measurements may be extracted from $\Delta
\textbf{L}_g[x,z]$. The quadratic average of $\Delta
\textbf{L}_g(\theta)$, denoted $Q(\theta)$, is a probe of the
amplitude of small relative grain movements (a fraction of the grain
size). Since this average also integrates the detector noise, we
evaluate the contribution of the latter, denoted $Q_0$, by measuring
the quadratic noise on successive pictures without rotation and
taken under the same conditions as in the experiment. We subtract
this value so that $Q(\theta)$ only reflects grains motions:

\begin{equation}\label{q}
Q(\theta) =  \left\langle \Delta
\textbf{L}_g(\theta)^2\right\rangle_{[x,z]} - Q_0
\end{equation}

We can also extract, for the same measurements, the displacement
$(\Delta x_b,\Delta z_b)$ of the barycenter of the pack, for every
angular increment, $\delta \theta$:

\begin{equation}\label{dxz}
    \left(\begin{array}{c} \Delta x_b(\theta) \\ \Delta z_b(\theta)
    \end{array} \right) = K \cdot \sum_{\{x,z\}} \left(\begin{array}{c} x \\ z
    \end{array} \right) \Delta \textbf{L}_g(\theta)[x, z]
\end{equation}

\noindent The sum is performed over all the positions $[x,z]$. $K$
is chosen in order to express the distances $\Delta x_b$ and $\Delta
z_b$ in drum radius unit. Though expressed as a length, $\Delta z$
may also be interpreted as an energy variation.

Figure \ref{figure1}(c) displays the evolution of the quadratic
fluctuations $Q(\theta)$. Its spikes are a direct signature of the
existence of intermittency in the micro-movements of the grains,
well before the surface reaches the avalanche angle. This
observation is consistent with the 2D numerical simulations of
Staron et al. \cite{Staron2002}. The discrete events become
detectable when the surface inclination approaches the horizontal.
We thereafter observe a progressive increase of the magnitude of
$Q(\theta)$, but no significant evolution of the intermittency
frequency.

\begin{figure}
\centerline{ \epsfxsize=8.5truecm \epsfbox{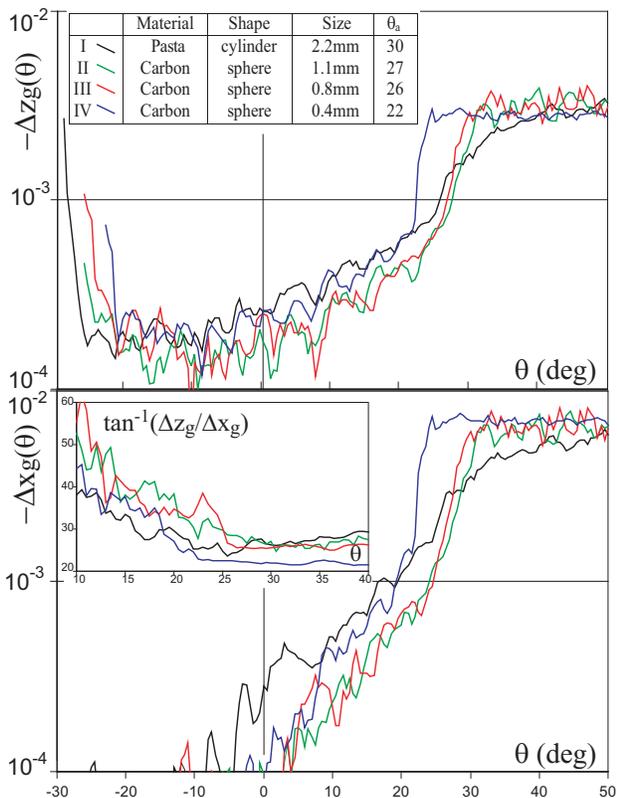}}
\caption{Normalized displacement of the pack barycenter for four
different systems. The length unit is set as the drum radius. (I)
is the average over 15 runs, whereas (II), (III) and (IV) are
averaged over 25 runs. The inset describes the angular direction
of the displacement.}
 \label{figure4}
 \end{figure}

The displacements of the pack barycenter, on different systems,
and with different grains sizes, are displayed in figure
\ref{figure4}. Three regimes may be distinguished. For
$-\theta_a<\theta<0$, we measure small values of $\Delta z_b$,
whereas $\Delta x_b$ remains insignificant. This indicates that a
slight compaction occurs in the sample as expected when a granular
system is mechanically perturbed. For $0<\theta<\theta_a$, we
observe a strong increase on both $\Delta x_b$ and $\Delta z_b$.
This evolution is associated with a progressive increase of the
mean displacement direction, given by $\tan^{-1}(\Delta x_b/\Delta
z_b)$, that eventually aligns with the free surface, figure
\ref{figure4} inset. As the inclination reaches the avalanche
angle $\theta_a$, we observe a sudden increase of the dynamics;
beyond $\theta_a$, the flow occurs along the surface of the pack
($\left\langle\tan^{-1}(\Delta x_b/\Delta z_b)\right\rangle =
\theta_a$). The values of $\Delta x_b$ and $\Delta z_b$ simply
correspond to the average grain flow necessary to compensate for
the rotation of the drum.

\begin{figure}
\centerline{ \epsfxsize=5.5truecm \epsfbox{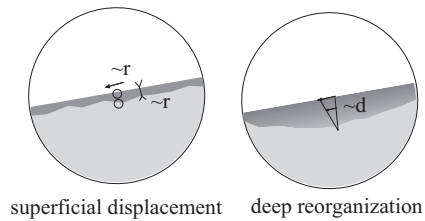}}
\caption{Illustration of the superficial and deep displacements.
Case 1 (left): the displacement involves a superficial layer which
thickness scales with the grain size $r$, moving over a distance
of the order of $r$. The barycentric displacement
$\Delta x$ then scales like $r^2$.\\
Case 2 (right): the failure corresponds to a minute deformation of
a layer whose thickness $d$ is controlled by the drum size: in
this case, the barycentric displacement is independent of the
grain size.} \label{sketch}
\end{figure}

Surprisingly, the measurement in the pre-avalanche regime seems to
be independent of the grain type, and in particular of the grain
size. This observation can be used to discriminate between two
possible interpretations (see figure \ref{sketch}): (a) If the
dynamics were due to a purely superficial effect, such as grains
rolling, the thickness of the moving layer and the magnitude of
the displacement would be controlled by the grain size. Hence, the
average displacement in the pre-avalanche regime, expressed in
units of drum radius, should scale as the square of the grain
size. This is inconsistent with our observations. (b) In contrast,
the independence of the dynamics with grain size strongly suggests
that the surface flow penetrates onto a depth $d$ controlled by
the drum geometry (diameter and/or thickness). The pre-avalanche
regime thus corresponds to a macroscopic instability of the pack
structure, though preferentially occurring in the vicinity of the
surface.

\begin{figure}
\centerline{ \epsfxsize=8.5truecm \epsfbox{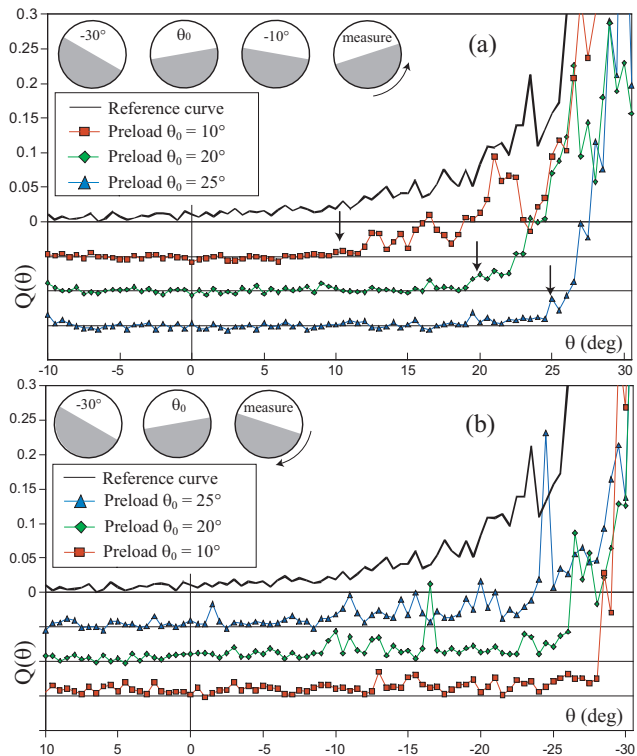}}
\caption{Role of the preparation (grain type I) (a) Effect of a
pre-load until an angle $\theta_0$ on the dynamics. (b) Evaluation
of the range of stability for different preparations. All the
curves correspond to the average of four similar runs (except 15
for the reference curve). For a better visibility, graphs have
been shifted.} \label{figure3}
\end{figure}

In considering figure \ref{figure4}, it is tempting to describe
this instability as being controlled by the surface angle. To
probe this hypothesis, we explore the effect of other parameters -
such as the preparation - on the pre-avalanche dynamics. The
following sequence of tilt is thus applied to the pack before
starting the measurement: the pack is first rotated $360^\circ$
clockwise, then slowly tilted backwards until the surface
inclination reaches a prescribed angle $\theta_0$, and finally
returned to a fixed value $-10^\circ$, figure \ref{figure3}(a).
From that point, the dynamics are recorded as the drum is rotated
counter-clockwise using the usual protocol.

The figure \ref{figure3}(a) shows the evolution of the quadratic
fluctuations $Q(\theta)$ (equation \ref{q}) obtained for three
different values of $\theta_0$, which are compared with the
reference curve obtained with data from figure
\ref{figure1}-inset. Interestingly, the dynamics are, in all
cases, negligible until the drum reaches the angle $\theta_0$.
Beyond this value, the activity increases rapidly and shows
similar features as in the non pre-loaded sample. This result
demonstrates that the pre-avalanche dynamics does not depend
solely on the surface inclination. The absence of detectable
plastic events from $-10$ to $\theta_0$ shows that the successive
static configurations that the system experiences during a tilt
experiment are stable over a large range of surface inclinations.

To probe in more detail the range of stability of each structure
we perform a complementary experiment depicted in Figure
\ref{figure3}(b). As previously shown, the system is rotated
$360^\circ$ clockwise, then tilted backwards to a prescribed angle
$\theta_0<\theta_a$. The dynamics is then recorded as the pack is
rotated clockwise towards the avalanche angle. We observe that the
threshold angle for which detectable motion occurs, and
consequently the range of mechanical stability of the pack, depend
on the preparation angle $\theta_0$. For $\theta_0$ of the order
(or less) than $0$, the pack remains stable until we reach
$-\theta_a$. For larger values of $\theta_0$, the onset of
internal dynamics occurs after a rotation of roughly $30^\circ$,
regardless of $\theta_0$. These results indicate that each
structural state of the pack is associated with a range of
inclination angle within which it is mechanically stable (within
the resolution of our measurements). The width of this stability
range barely depends on the pre-charge angle, but the latter
controls the
value of its limiting angles.\\

From the observed pre-avalanche dynamics, one might be tempted to
extrapolate information about the triggering of macroscopic
avalanche flows such as the avalanche angle. However, we find that
grains with different avalanche angles exhibit quantitatively
similar pre-avalanche behaviors, figure \ref{figure4}. Moreover,
we show that the preparation and history of the pile have a major
impact on the statistics of the micro-events, but no significant
effect on the avalanche angle \cite{Grasselli}. As a consequence,
it seems unlikely that the measure of the pre-avalanche dynamics
may serve as a tool for a robust prediction of the macroscopic
avalanche.

These observations might reflect the fact that the two
instabilities - pre-avalanche failures and avalanche triggering -
are qualitatively different. The discrete micro-failures we have
observed correspond to the mechanical destabilization of a static
structure induced by an incremental tilt. The energy released by
each event is instantaneously dissipated. The instability is thus
only controlled by the static properties of the granular
structure. In contrast, the avalanche is a dynamical instability
that primarily involves the dissipative properties of the
material. An avalanche is triggered when the pile cannot dissipate
the inertial energy produced by a any small failure
\cite{Daerr1999,Bouchaud1995,Rajchenbach2002}.

We have presented here a new experimental technique suitable for
the study of 3D confined granular dynamics, that allowed us to
measure minute displacements of the structure. In contrast to
other techniques (DWS \cite{Kabla2004} capacitive measurements
\cite{Josserand2000}), our method allows for local and
quantitative measurements of the deformation. We are currently
using X-ray computerized tomography in order to obtain 3D
information on these plastic events \cite{Aste2004}. Beyond the
displacement field, this method might enable us to directly probe
the statistical properties of the contact network, which might
control the plastic response of the material.

\begin{acknowledgments}

We wish to thank Tomaso Aste for stimulating discussions, and Tim
Sawkins for his helpful technical contributions.

\end{acknowledgments}

\end{document}